# Twin Transition or Competing Interests?

Validation of the Artificial Intelligence and Sustainability Perceptions Inventory (AISPI)

Annika Bush
Research Center Trustworthy Data Science and Security of the University Alliance Ruhr, Faculty of Computer Science, Technical University Dortmund
Dortmund, Germany
annika.bush@tu-dortmund.de

## Abstract

As artificial intelligence (AI) and sustainability initiatives increasingly intersect, understanding public perceptions of their relationship becomes crucial for successful implementation. However, no validated instrument exists to measure these specific perceptions. This paper presents the development and validation of the Artificial Intelligence and Sustainability Perceptions Inventory (AISPI), a novel 13-item instrument measuring how individuals view the relationship between AI advancement and environmental sustainability. Through factor analysis (N = 105), we identified two distinct dimensions: 'Twin Transition' and 'Competing Interests'. The instrument demonstrated strong reliability ($\alpha$ = .89) and construct validity through correlations with established measures of AI and sustainability attitudes. Our findings suggest that individuals can simultaneously recognize both synergies and tensions in the AI-sustainability relationship, offering important implications for researchers and practitioners working at this critical intersection. This work provides a foundational tool for future research on public perceptions of AI's role in sustainable development.

## 1 Introduction

The rapid advancement of artificial intelligence (AI) is reshaping industries, societies, and our approach to global challenges [2, 19]. As concerns about climate change, injustice, peace, and environmental degradation intensify, understanding how AI can contribute to or potentially hinder sustainable development has become a central question [1, 8]. Public perception of this relationship is crucial in shaping policy, adoption, and implementation of AI-driven sustainability solutions [18]. This research addresses this methodological gap by developing and validating the Artificial Intelligence and Sustainability Perceptions Inventory (AISPI), a new survey instrument designed to measure these specific perceptions. Through rigorous psychometric validation, the aim is to provide researchers and practitioners with a reliable tool for assessing how individuals view the relationship between AI advancement and environmental sustainability.

*Research Question:* What are the psychometric properties and underlying factor structure of the Artificial Intelligence and Sustainability Perceptions Inventory (AISPI) for measuring public perceptions of the interrelatedness of AI and sustainable development?

## 2 Related Work

### 2.1 Public Perceptions of AI

Studies of AI perceptions reveal complex and often contradictory attitudes while people increasingly recognize AI's role in technological and societal progress. Eubanks [7] and Osasona et al. [20] document persistent concerns about ethical implications, including privacy risks and decision-making transparency. In 2024, the 3M State of Science Index showed that 75% of people globally agreed that AI is an exciting technology. 80% even agreed that AI can help build a more sustainable future [21].

Several studies documented that young adults' views on AI are significantly influenced by personal experience and knowledge level, suggesting that technological familiarity does not automatically translate to uncritical acceptance [15, 16]. Recent research has shown that public perceptions of AI are multifaceted and deeply rooted in social identity and cultural dimensions [13]. Jensen et al. [13] found that ideas about humanness and ethics are central to how experts and the general public perceive AI. The public's views are often shaped by comparing AI's characteristics to human capabilities, particularly concerning empathy, nuance handling, and agency. Studies indicate that perceptions are influenced by fundamental values, social values, economic standings, and demographics [14]. People's trust in AI is contextual - they may appreciate its efficiency in routine tasks but remain skeptical about its role in complex situations requiring human judgment. The 3M State of Science Survey [21] revealed a paradoxical nature in public perception, where AI is seen simultaneously as an opportunity and a potential threat requiring regulation.

## 2.2 Society and Sustainable Development

The United Nations' Sustainable Development Goals (UN SDGs) [25], adopted in 2015, represent a global framework of 17 interconnected goals addressing humanity's major challenges. These goals span environmental protection (e.g., climate action, life below water), economic development (e.g., decent work and economic growth), and social progress (e.g., quality education, reduced inequalities). While these goals are often categorized into ecological, economic, and social dimensions, the UN emphasizes their interconnected and indivisible nature [25].

Since their adoption, the SDGs have helped shape public discourse and understanding of sustainability challenges, particularly among younger generations. This growing awareness is reflected in recent surveys and social movements. As noted by Seth (2024), "Climate change and environmental sustainability related topics feature heavily amongst the problems to solve that have been top of mind for the global public" [21, p. 25].

The current generation of young adults demonstrates unprecedented environmental awareness and commitment to sustainability. Li et al. [17] document stronger pro-environmental attitudes compared to previous generations, while Hamid and Al Mubarak [11] highlight how responsibility toward future generations motivates interest in sustainable technologies. Social media movements like #FridaysForFuture have amplified youth engagement with environmental justice [5, 12], potentially influencing perceptions of technological solutions to sustainability challenges.

## 2.3 AI-Sustainability Intersection

The intersection of AI and sustainability has emerged as a critical area of study, highlighting both transformative opportunities and significant challenges. Research indicates that AI holds substantial potential for advancing sustainability goals through environmental monitoring, climate prediction, and resource optimization. Vinuesa et al. [26] demonstrate AI's capacity to support UN SDGs through improved data-driven decision-making and resource optimization. Experts broadly agree that AI can positively contribute to these goals, particularly in climate action and environmental protection.

However, this relationship is complex and sometimes paradoxical. While AI can enhance efficiency in resource management and environmental monitoring, important concerns persist about its own environmental footprint. Strubell et al. [23] specifically highlight the substantial energy demands of large-scale machine learning models, raising questions about the net environmental impact of AI deployment. This tension between AI's potential benefits and its environmental costs remains understudied from a public perception perspective.

Recent research has highlighted both the opportunities and challenges in applying AI towards UN SDGs. While AI shows promise for addressing social challenges, successful implementation requires careful consideration of multiple factors. Studies have shown the uneven distribution of AI applications across different SDGs [6] and identified key requirements for successful collaborations between AI researchers and domain experts [24]. These findings emphasize the importance of a systematic approach that considers both the technical potential of AI solutions and the practical realities of implementation in social contexts. The need to evaluate whether AI and sustainability initiatives are viewed as competing or complementary transitions by different stakeholders is particularly relevant, as this perception can significantly impact adoption and effectiveness.

Recent surveys provide insight into public attitudes toward this intersection. The 2023 3M State of Science survey [21] found that 75% of global respondents believe "green jobs" are crucial for addressing climate change, with 89% agreeing that science and technological innovation should advance planetary well-being. These findings suggest broad public support for technological solutions to environmental challenges while also highlighting the need to carefully consider the environmental impact of AI systems.

The relationship between AI and sustainability is multifaceted - AI can serve as a tool for environmental solutions, acts as a direct contributor to environmental challenges through its resource demands, and may fundamentally challenge sustainability goals through its societal and economic impacts. Understanding how the public perceives these complex interactions becomes crucial for developing and implementing AI solutions that are environmentally responsible, socially accepted, and aligned with broader sustainability objectives.

## 2.4 Research Gap

While validated instruments exist for measuring public attitudes toward AI [9, 22] and sustainable development [3, 27] separately, there is currently no psychometrically validated tool for assessing how individuals perceive the relationship between AI advancement and sustainability goals. This measurement gap is particularly critical given the complex and sometimes paradoxical relationship between AI and environmental sustainability highlighted in recent literature [23, 26].

Existing research suggests that public perceptions of both AI and sustainability are multifaceted and influenced by various factors, including personal experience, cultural dimensions, and social identity [13, 14]. However, the lack of a validated measurement instrument has limited our ability to systematically assess how individuals conceptualize the relationship between AI advancement and sustainability efforts. This gap hampers both research and practice, as understanding these perceptions is crucial for developing and implementing environmentally responsible AI solutions that gain public acceptance.

The methodological gap was addressed by developing and validating the AISPI. This instrument measures explicitly how individuals perceive the potential synergies and tensions between AI advancement and sustainable development, providing researchers with a reliable tool for future investigations in this critical area.

## 3 Method

This study employed a mixed-methods approach, as recommended by Boateng et al. [4], to validate the new survey instrument measuring perceptions of AI's role in sustainable development. The validation study combined our newly developed scales with existing validated instruments to ensure a comprehensive assessment of the construct. The initial survey version was given to six participants who gave qualitative feedback on its items and their comprehensibility. Afterwards, the survey was improved and quantitatively tested.

## 3.1 Survey Design

Our validation study combined newly developed scales with established instruments in a comprehensive survey. The core component was our newly developed AI-Sustainability Perceptions Inventory (AISPI), which was supplemented with validated measures to assess construct validity, including both convergent and discriminant validity.

*3.1.1 Existing Instruments.* For measuring AI attitudes, the four-item AIAS-4 scale [9] with a 6-point Likert scale (1 = "Strongly Disagree" to 6 = "Totally Agree") was used. To assess sustainability attitudes, the 11-item Sustainability Attitude Scale (SAS) [26] was modified after initial pilot testing revealed social desirability concerns by the participants. The adaptation included inverting six items to reduce acquiescence bias while maintaining the scale's core construct measurement.

*3.1.2 AI-Sustainability Perceptions Inventory (AISPI).* The AISPI was developed systematically following established scale development guidelines [4]. Initial item generation was informed by existing literature on AI perceptions and sustainability attitudes, resulting in an initial pool of 19 items. These items were refined through expert review and interviews with six participants to ensure content validity and item clarity. Our refined version for the validation study consisted of 14 items measured on a 6-point Likert scale (1 = "Strongly disagree" to 6 = "Totally agree").

Items were designed to capture different aspects of the perceived relationship between AI advancement and sustainability efforts. Example items include "AI and sustainability efforts can be mutually reinforcing," "Sustainable development will limit the development of AI," and "The energy consumption of AI systems hinders sustainability efforts." The 6-point scale was chosen to eliminate the neutral midpoint and encourage participants to express a directional opinion.

*3.1.3 Participants.* Participants were recruited through private and professional networks as well as social media. They were required to be over 18 years old and proficient in English. To ensure diverse perspectives, people from all over the world were invited to participate. Participation was voluntary and not compensated.

## 4 Results

## 4.1 Demographics

The final sample (N=105) included participants from 12 different countries, with the largest representation from Germany (73.3%), followed by the UK (3.8%) and USA (3.8%). Participants' ages ranged from 18 to >65 years, with most participants being 25-34 years old (32.4%).

## 4.2 Reliability Analysis

The internal consistency of all measurement scales was assessed using Cronbach's alpha (see Tab. 1). The adapted Sustainability Attitudes Scale (SAS) [27] showed excellent reliability ($\alpha$ = .90, 12 items). The Artificial Intelligence Attitudes Scale (AIAS-4) [9] demonstrated very high internal consistency ($\alpha$ = .93, 4 items). The newly developed scales also showed strong reliability: the AISPI scale ($\alpha$ = .90, 13 items) and the newly developed AI and SDG-17 scale ($\alpha$ = .95, 17 items) both exceeded recommended thresholds for research purposes.

The internal consistency of all measurement scales was assessed using Cronbach's alpha (see tab. 1). The adapted Sustainability Attitudes Scale (SAS) [27] showed excellent reliability ($\alpha$ = .90, 12 items). The Artificial Intelligence Attitudes Scale (AIAS-4) [9] demonstrated very high internal consistency ($\alpha$ = .93, 4 items). The newly developed scales also showed strong reliability: the AISPI scale ($\alpha$ = .90, 13 items) and the newly developed AI and SDG-17 scale ($\alpha$ = .95, 17 items) both exceeded recommended thresholds for research purposes.

**Table 1: Internal consistency of all used measurement scales**

| Name of Instrument | Items | Cronbach's Alpha |
|---|---|---|
| SAS [27] (adapted version) | 11 | 0.90 |
| AIAS-4 [9] | 4 | 0.93 |
| AISPI (new) | 13 | 0.89 |

## 4.3 Factor Structure and Inter-Factor Relationship

Preliminary factor analysis revealed that one item ("Sustainable development is more important than AI") showed poor factor loading (< .40) and did not align well with either emergent factor. After removing this item, the final 13-item scale demonstrated high internal reliability ($\alpha$ = .89) while achieving a clearer factor structure.

The factor structure of the AISPI was examined using exploratory factor analysis (EFA). The Kaiser-Meyer-Olkin measure verified the sampling adequacy for the analysis, KMO = .891, which is well above the acceptable limit of .5. Bartlett's test of sphericity, $\chi^2(78) = 709.845, p < .001$, indicated that correlations between items were sufficiently large for EFA.

Principal axis factoring was conducted on the 13 items with oblique rotation (Oblimin). Two factors emerged based on the Kaiser criterion of eigenvalues greater than 1. The two-factor solution explained 52.16% of the variance. Factor 1, representing the 'Twin Transition' of both AI and sustainability, accounts for 40.68% of the variance. Factor 2 represents 'Competing Interests' of AI and sustainable development, explaining an additional 11.48% of the variance. This level of explained variance meets common thresholds in social science research, where 50-60% of total explained variance is typically considered acceptable [10].

The pattern matrix showed clear loadings on two distinct factors (see 1). Factor 1 comprised eight items with loadings ranging from .510 to .887, reflecting positive perceptions of AI's role in sustainability and vice versa. The highest loading items included "AI can help optimize resource use and reduce waste" (.887) and "AI will advance sustainable development" (.830). Factor 2 contained five items with loadings ranging from .574 to .864, representing perceived conflicts between AI and sustainability efforts. The strongest loading items were "AI will hinder sustainable development" (.864) and "Sustainable development will limit AI advancements" (.804).

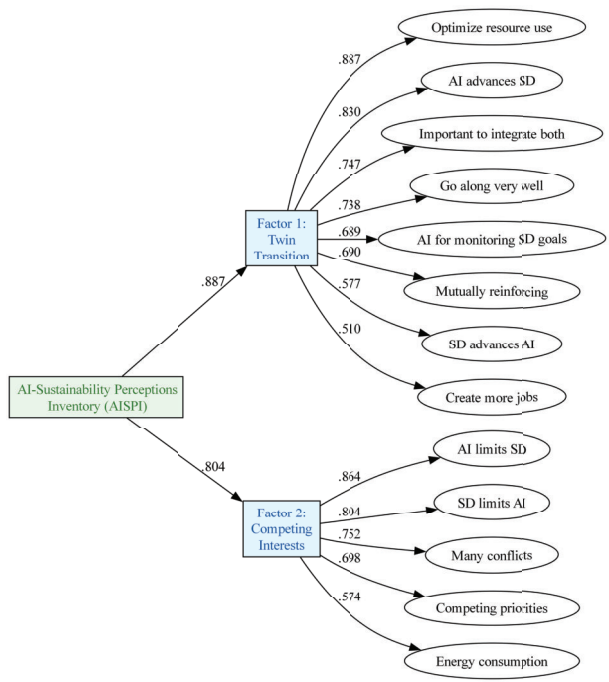

**Figure 1: Two-factor structure of the AI-Sustainability Perception Inventory with standardized factor loadings.**

Correlation analyses were conducted, and mean ratings were compared to examine the relationship between the identified factors. Analysis of the two AISPI factors revealed that participants rated the Twin Transition factor (Factor 1) slightly higher (M = 4.26, SD = 1.26) than the Competing Interests factor (Factor 2; M = 4.03, SD = 1.42). A Pearson correlation analysis showed a moderate positive correlation between the two factors (r = .488, p < .001, N = 105), indicating that higher ratings on the Twin Transition factor were associated with higher ratings on the Competing Interests factor. The correlation was statistically significant at the .01 level (two-tailed).

### 4.4 Convergent Validity

To establish the convergent validity of the AISPI, the correlations between its factors and related constructs were examined. Factor 1 ('Twin Transition') showed significant positive correlations with both AI attitudes (AIAS-4; r = .531, p < .001) and sustainability attitudes (SAS; r = .387, p < .001), supporting its construct validity. This suggests that participants who view AI and sustainability as a twin transition also tend to have more positive attitudes toward both AI and sustainability in general.

Factor 2 ('Competing Interests') showed a different pattern, correlating moderately with sustainability attitudes (SAS; r = .264, p = .007) but showing no significant correlation with AI attitudes (AIAS-4; r = .055, p = .578). This distinct pattern of correlations supports the discriminant validity of our two-factor structure, suggesting that the perception of AI-sustainability conflicts represents a unique construct that is not merely a reflection of general AI attitudes.

## 5 Discussion

The validation of the AI-Sustainability Perceptions Inventory (AISPI) represents an important methodological contribution to understanding how individuals perceive the relationship between AI advancement and sustainability efforts. Our findings reveal several key insights about the instrument's psychometric properties and theoretical implications.

The emergence of two distinct factors – 'Twin Transition' and 'Competing Interests' - demonstrates that individuals hold nuanced views about the relationship between AI and sustainability. This complexity mirrors recent findings in the literature regarding the paradoxical nature of AI's role in sustainability efforts[23, 26], where AI can simultaneously serve as both a tool for environmental solutions and a contributor to environmental challenges through its resource demands.

The strong factor loadings and acceptable explained variance (52.16%) support the construct validity of the AISPI, while the high internal consistency ($\alpha$ = .89) indicates that the instrument reliably measures these perceptions. The clear two-factor structure suggests that people can simultaneously recognize both synergies and tensions in the AI-sustainability relationship. This is further supported by the moderate positive correlation (r = 0.488, p < .001) between Twin Transition and Competing Interests factors, indicating that participants who strongly recognized the synergistic potential of AI and sustainability were also more likely to acknowledge potential conflicts between these domains. The strong reliability of both the adapted SAS ($\alpha$ = .90) and AIAS-4 ($\alpha$ = .93) in our study further validates their integration alongside the AISPI for a comprehensive assessment of attitudes toward AI and sustainability.

The two-factor structure of the AISPI advances our theoretical understanding of how individuals conceptualize the relationship between technological advancement and environmental sustainability. The simultaneous recognition of both synergies and tensions suggests a sophisticated cognitive framework that goes beyond simple positive or negative attitudes. This finding aligns with recent research by Jensen et al. [13] showing that public perceptions of AI are multifaceted and deeply rooted in broader social contexts. Similar to how Kanzola et al. [14] found that AI perceptions are influenced by fundamental values and social factors, our findings suggest that people's understanding of AI's role in sustainability is equally complex.

The moderate positive correlation between factors suggests that increased awareness of AI-sustainability relationships may lead to more nuanced perspectives rather than polarized views. This finding resonates with recent survey data [21] showing that people can simultaneously view AI as both an opportunity and a potential challenge requiring careful consideration, particularly in the context of environmental sustainability. This has important implications for how we understand cognitive frameworks around technology adoption and environmental attitudes. Our findings suggest a complex interplay of perceptions that must be considered when studying AI implementation in sustainability contexts.

While adequate for initial validation, our sample size (N=105) suggests the need for larger-scale validation studies across different cultural contexts and demographic groups. This is particularly important given that recent research [13, 14] has demonstrated how cultural dimensions and social identity significantly influence AI perceptions. To address this limitation, the survey will be translated into several languages and conducted as a large-scale study to facilitate a comprehensive international assessment of AI-sustainability perceptions.

## 6 Conclusion

The validation of the AISPI represents a significant step forward in understanding public perceptions of AI's role in sustainability efforts. The instrument's robust psychometric properties and factor structure provide researchers and practitioners with a valuable tool for assessing these complex perceptions. Through this validation study, it has been demonstrated that the AISPI reliably captures the multifaceted nature of how individuals view the relationship between AI advancement and sustainability efforts.

The successful validation of the AISPI addresses a crucial gap in the literature by providing researchers with a reliable instrument for measuring these specific perceptions. The emergence of a two-factor structure suggests that future research should consider both potential synergies and conflicts when examining public attitudes toward AI in sustainability contexts. This nuanced understanding is particularly important as organizations and policymakers work to implement AI solutions for environmental challenges.

The AISPI can serve as a valuable tool for researchers, policymakers, and practitioners working at the intersection of AI and sustainability. For researchers, it provides a validated instrument for investigating how public perceptions influence the acceptance and implementation of AI-driven sustainability initiatives. Policymakers can use the instrument to better understand public concerns and support for various AI-sustainability policies, enabling more effective communication strategies and evidence-based policy development. Practitioners in both AI and sustainability fields can benefit from insights into how their work is perceived and what factors might influence public support or resistance.

## A Survey items

Artificial Intelligence and Sustainability Perceptions Inventory (AISPI). For this study, is has been used with a 6-point Likert scale (1 = strongly disagree, 6 = strongly agree)

(1) AI can help optimize resource use and reduce waste.
(2) AI will create more jobs than it will eliminate.
(3) The energy consumption of AI systems could hinder sustainability efforts.
(4) AI is essential for monitoring and achieving sustainability goals.
(5) The pursuit of AI advancement and sustainability are competing priorities.
(6) AI and sustainability efforts can be mutually reinforcing.
(7) Sustainable development will limit AI advancements.
(8) There are many conflicts between the advancement of AI and sustainability efforts.
(9) AI will hinder sustainable development.
(10) AI will advance sustainable development.
(11) AI and sustainable development go along very well.
(12) It is important for society to integrate both AI advancement and sustainability efforts.
(13) Sustainable development will advance the development of AI.